\documentstyle[aps]{revtex}
\let\institute=\address
\begin{document}

\title{\vspace*{-35pt}{\small{\it Europhys. Lett.,} {\bf 43}(6), p.p. 623-628 (1998)}\\[10pt] 
Magnetic Force Exerted by the Aharonov-Bohm Line
}
\author{A.L. Shelankov $^{*}$}
 \institute{ Department of
    Theoretical Physics, Ume{\aa} University, 901 87 Ume{\aa}, Sweden
    }

\maketitle

\begin{abstract}
The controversial question of the transverse force exerted by the
Aharonov-Bohm (AB) magnetic flux line is reconsidered with the help of
a new paraxial solution to the AB-scattering problem.  It is shown
that despite the left-right symmetry in the AB scattering
cross-section, a beam of a finite width is deflected by the AB-line as
if by the ``Lorentz'' force.  The asymmetry and the magnetic force
originate from the quantum interference in the forward direction
within the angular size of the incident wave.  In the context of the
superfluid He-4, the paper confirms the Iordanskii force acting on the
vortex line.  
\end{abstract}

\vspace{15pt}

The Aharonov-Bohm (AB) flux line is an idealized construction
originally designed to discuss the role of the vector potential in
quantum mechanics \cite{AhaBoh59,OlaPop85,PesTon86}.  The magnetic
field around the AB-line is zero but the gauge vector potential is
finite being generated by the magnetic flux $\Phi $ concentrated in
the line.  Nowadays, the AB-line concept finds its application in
various contexts: As a carrier of Chern-Simon's field, the AB-line
attached to a particle allows one to build two-dimensional objects
obeying fractional statistics (anyons), or composite fermions in the
theory of strongly correlated electronic systems (Quantum Hall Effect
or high-T$_{c}$ oxides).  In many respects, vortex lines in
superfluids are similar to AB-lines: The superflow around a vortex
line in $He-II$, although not a gauge field, acts on a normal
excitation (with some reservations) like the vector potential of an
AB-line on a charge \cite{Son75,Son97,DavSte97,WexTho98}.  Relevant in
many contexts, the AB-problem has been the subject of enormous amount
of papers (for a review see \cite{OlaPop85,PesTon86}).

Surprisingly, there is a question which is still controversial: Does
the AB-line exert a Lorentz-like magnetic force on a moving charge?
Or, in other words, whether the charge is deflected right-left
asymmetrically revealing the absence of the mirror symmetry (parity
${\cal P}$) broken by the magnetic flux.  Despite the fact that the
exact solution to the AB-line scattering problem has been long known
\cite{AhaBoh59,OlaPop85,PesTon86}, there is no unanimity about the
subject.  In the context of superfluidity, a closely related question
is about the origin and the very existence of the Iordanskii force
\cite{Ior65} -- the force acting on the vortex from the normal
component, transverse to the direction of their relative motion.  This
question is in the centre of recent debates in the literature on
vortex dynamics initiated by \cite{Magnusnia97} and continued in
\cite{Son97,WexTho98,ThoAoNiu97}.

The existing controversy can be explained as follows.  It seems to be
a common opinion that the momentum transfer can be expressed via the
differential cross-section $d \sigma / d \varphi = |f|^{2}$,
$f(\varphi)$ being the scattering amplitude.  Then, the transverse
force is proportional to $\sigma_{\perp}$,
\begin{equation}
\sigma_{\perp} =
\int_{- \pi }^{\pi } d\varphi \; \sin \varphi \;|f(\varphi )|^{2}\;\; ,
\;\;  
|f^{(AB)}(\varphi )|^{2}
=
{\lambdabar  \over 2 \pi}\;
{\sin^{2}\pi \tilde{\Phi} \over \sin^{2}{\varphi \over 2 } } 
\;\; ,
\label{gma}
\end{equation}
$f^{(AB)}$ is the AB-scattering amplitude \cite{AhaBoh59} ($\lambdabar
=1/k$, $\tilde{\Phi}= \Phi/\Phi_{0},\; \Phi_{0}= hc/e $).  Although
the mirror transformation $\varphi \rightarrow -\varphi $ does change
the phase of $f^{(AB)}$ the cross-section $|f^{(AB)}|^{2}$ is
left-right {\it symmetric}.  If so, the particle acquires momenta in
the right and left direction with equal probability, and the net
transverse momentum transfer vanishes.

On the other hand, the scattering amplitude can be presented as a sum
of partial waves, $f(\varphi )= \Sigma_{m}f_{m}\exp (im\varphi )$, and
$\sigma_{\perp}$ in Eq.(\ref{gma}) apparently identically transforms
to,
\begin{equation}
\sigma_{\perp}= \lambdabar 
\sum_{m= - \infty }^{\infty } \sin 2(\delta_{m}-\delta_{m+1})
\;\; , \;\;  
\delta_{m}^{(AB)}= {\pi \over 2}\big(|m| - |m- \tilde{\Phi}| \big)
\label{lma}
\end{equation}
where $\delta_{m}$ is the partial wave phase shift, and
$\delta^{(AB)}_{m}$ is the AB phase shift \cite{Hen80}.  Only one term
in the sum is nonzero, and $\sigma^{(AB)}_{\perp}=- \lambdabar \sin
2\pi \tilde{\Phi}$.  Compatible with the broken parity, this {\it
finite} value of the transverse force has been found by several
authors in the context of superfluidity theory
\cite{Ior65,Son75,Son97} and the magnetic line problem
\cite{OlaPop85,Cle68,NieHed95}.  Recently Eq.(\ref{lma}) as well as
conclusions based upon it, have been claimed wrong
\cite{ThoAoNiu97,WexTho98} by the argument that its derivation from
Eq.(\ref{gma}) includes dubious manipulations with diverging sums.
The purpose of the present paper is to reconsider the problem and to
resolve the contradiction.

Clearly, the difficulties are due to a singularity in the forward
scattering, which in turn is due to the infinite range of the vector
potential of the AB-line. As shown below, the standard approach fails
if the leading contribution comes from forward scattering: Derived
classically, Eq.(\ref{gma}) is incomplete -- it misses a quantum
interference term which is responsible for the finite transverse
force.  Therefore, it is true that the two representations for
$\sigma_{\perp}$ are not equivalent \cite{ThoAoNiu97,WexTho98} in the
case of AB-scattering.  The correct one is that in Eq.(\ref{lma}), as the
present paper asserts.

A natural way of analysing the forward scattering singularity is to
scatter wave packets (beams, in a stationary theory) rather than plane
waves.  The solution corresponding to a beam-like incident wave may be
built by superimposing the Aharonov-Bohm wave functions as in
\cite{Ste95Alv96}.  For our purposes, it is more practical to use a
simplified version of the Schr\"odinger equation -- the paraxial
(parabolic) approximation \cite{parax} -- valid for the description of
the small angle scattering.  At the expense of fine details on the
scale of the wave length $\lambdabar$, the paraxial approximation
allows one to find rather easily the wave function for small
scattering angles.

The paper is organized as follows.  First, the paraxial approximation
is discussed, the solution for a general incident wave is found and
compared with the known AB-solution.  Then, to check the very
existence of the left-right asymmetry in scattering, a gedanken
experiment is analysed, where a beam hits an AB-line and the
deflection of the beam as a whole measures the transverse ``Lorentz
force'' exerted by the line.  As expected from the symmetry arguments,
a finite deflection is observed.  The transverse force is found for
arbitrary incoming wave.  The physics behind the inequivalence of
Eqs.(\ref{gma}) and (\ref{lma}) is discussed using the S-matrix
formalism.  The results are summarised at the end of the paper.

In the paraxial theory \cite{parax}, a particle moving on the $x-y$
plane at a small angle to the $x-$axis is described by the wave
function of the form $\Psi = e^{ikx} \psi$, where $\psi (x,y) $ is a
slowly varying envelope, $ |\nabla \psi | \ll k |\psi|$.  Neglecting
${\partial^{2} \psi\over{\partial x^{2}}}$ $\ll k {\partial \psi
\over{\partial x}}$ in the stationary Schr\"odinger equation, one
comes to the paraxial equation:


\begin{equation}
i  v \partial_{x}\psi  = -{1\over{2m}}\partial^{2}_{y}\psi
\label{4ea}
\end{equation}
where the velocity $v = \hbar k/m $ and $\bbox{\partial}\equiv \hbar
\bbox{\nabla }-i {e\over{c}}\bbox{A}$; the vector potential $\bbox{A}$
of the AB-line piercing the plane at $\bbox{r}=0$ is conveniently
chosen as $A_{y}=0$ and $A_{x}= - \Phi \delta (x) \, y/2|y|$.

The incident wave coming from $x<0$, $\psi(x<0, y)$, is controlled by
conditions of the experiment such as screens, apertures {\em etc}.
Leaving the preparation of the incoming wave out of the picture,
$\psi_{in}(y)\equiv \psi(x= - 0, y)$ can be taken as the input to the
scattering problem.  In the immediate vicinity of the line $x=0$,
$\exp [(e/i \hbar c)\int_{-0}^{x}A_{x}dx]\psi_{in} $ solves
Eq.(\ref{4ea}), and one finds%
\begin{equation}
\psi(+0,y)= \psi_{in}(y)\exp \left(-\pi i \tilde{\Phi }\,
\hat{y}\right) 
\;\; , \;\;  \hat{y}\equiv y/|y|
\label{5ea}
\end{equation}
Further propagation is free, and the outgoing wave is
\begin{equation}
\psi(x>0,y)= \int_{- \infty }^{\infty }
d y'\,
G_{0}(y-y';x)
\psi(+0,y') \;\; , \;\;  
G_{0}(y;x)=  \theta (x)
{1\over \sqrt{ 2\pi ix \lambdabar}}
 e^{{iy^{2}\over{2  x \lambdabar}} } .
\label{55ea}
\end{equation}
Substituting $\psi(+0,y)$  Eq.(\ref{5ea}),
\begin{equation}
\psi (x>0,y)      =    
\cos \pi \tilde{\Phi}\,\,\psi_{0}(x,y)   
  + i \sin \pi \tilde{\Phi} 
\int_{- \infty }^{\infty }
d y'\,
G_{0}(y-y';x) 
\,\hat{y}'\,
\psi_{in}(y') 
\; ,
\label{hfa}
\end{equation}
where $\psi_{0}(x,y)= 
\int_{- \infty }^{\infty }
d y'\,
G_{0}(y-y';x)
\psi_{in}(y')
$
is the solution in the absence of the line.

This completes the derivation of the paraxial solution to the AB-line
scattering problem: Given the input $\psi_{in}(y)$, Eq.(\ref{55ea}) or
Eq.(\ref{hfa}) allows one to find the outgoing wave at $x \gg
\lambdabar $ at small angles $|\varphi | \ll 1$, $\varphi = y/x$.

If the incident wave is an {\em infinite} plane wave, {\it i.e.}
$\psi_{in}(y)=\psi_{0}(x,y)=1$, Eq.(\ref{hfa}) immediately gives (up
to a gauge transformation) the Aharonov-Bohm solution at $|\varphi
|\ll 1$ \cite{OlaPop85}, thus confirming the validity of the paraxial
approximation.  The paraxial solution $\psi = \psi (s)$ is a function
of $s = y/ \sqrt{2\lambdabar x}= \varphi \sqrt{x/2 \lambdabar}\,$.
\footnote{ $\psi (s)$ defines a curve on the complex $\psi$-plane. The
curve is the well-known Cornu spiral \cite{BorWol59}, the two centers
of which are at $\psi = \exp(\pm i \pi \tilde{\Phi})$; For $y$ varying
from $-\infty $ to $\infty $, $\psi (x,y)$ changes along the spiral
from $\psi = \exp( - i \pi \tilde{\Phi})$ to $\psi = \exp(i \pi
\tilde{\Phi})$.}  At $|s|\gg 1$, {\it i.e.}  $|\varphi| \gg
\sqrt{\lambdabar /x}$, the wave function acquires the asymptotic form
usual for a scattering problem.  Note, however, that at $y=s=0$ the
second term on the r.h.s. of Eq.(\ref{hfa}) vanishes, and $\psi=
\cos\pi \tilde{\Phi}$ at {\it any} distance from the AB-line
\cite{OlaPop85,BerChaLar80}.  The forward direction anomaly (when the
wave function differs from the incident wave and at the same time does
not depend on the distance to the scatterer) takes place at
$|s|\lesssim 1$ {\it i.e.}  in the progressively narrow angle range
$|\varphi| \lesssim \sqrt{\lambdabar /x}$. This is the singularity
which causes the calculational difficulties.

The $\varphi =0$ singularity is removed if the incident wave has a
finite width: From Eq.(\ref{55ea}), the beam $\psi_{in}(y)= \exp ( -
|y|/W )$ generates at $ x \gg W^{2}/\lambdabar$ the outgoing {\it
spherical wave} with the following angular distribution of the
intensity, $P(\varphi ) d \varphi = x |\psi|^{2} d \varphi $:
\begin{equation}
P(\varphi )= 
{2 \lambdabar  \over \pi   } 
{(\varphi \,\sin \pi \tilde{\Phi} -\varphi_{0}\cos \pi \tilde{\Phi})^{2}\over (\varphi^{2}+ \varphi_{0}^{2})^{2}}
\;\;
, \;\;  x \gg W^{2}/\lambdabar  \; ,
\label{ffa}
\end{equation}
$\varphi_{0} = \lambdabar / W$ 
being the beam angular width.  As expected, the angular distribution
is regular.  The AB cross-section is recovered at $|\varphi |\gg
\varphi_{0}$, whereas at $|\varphi |\lesssim \varphi_{0}$, the
distribution is {\em asymmetric} indicating a finite transverse
momentum transfer.  To quantify the asymmetry, the deflection of the
beam as a whole (``the trajectory bending'') is calculated below for a
general profile of the beam.

Consider the experiment where a particle moves on the $x-y$ plane in
the $x$-direction from $- \infty$ and meets the AB-line at
$\bbox{r}=0$.  The distribution with respect to the transverse
coordinate $y$ is measured, and the expectation value $\bar{y} (x)$
defines the ``trajectory'' from which the deflection of the particle
by the line is extracted.  To make the transverse coordinate
meaningful, the stationary incident wave is beam-like with a finite
transverse size $W\gg \lambdabar $.

The transverse position of the particle at a given $x$ is defined as $
\bar{y} = \int_{- \infty }^{\infty } dy\, y |\psi(x,y)|^{2}$, and the
angle of propagation is $\bar{\varphi} =d \bar{y}/ dx$.  It follows
from Eq.(\ref{4ea}) that $d \bar{y}/ dx = \langle
\hat{p}_{y}\rangle/mv$, $ \hat{p}_{y}$ being the kinematical momentum.
In the chosen gauge, $\langle \hat{p}_{y}\rangle_{out}= \int_{-\infty
}^{\infty }dy \, \psi^{*}(x,y){\hbar \over i} {\partial\over{\partial
y}}\psi(x,y)$ with $\psi (x,y)$ given by Eq.(\ref{55ea}).  In the free
motion region, $\langle \hat{p}_{y}\rangle_{out}$ does not depend on
$x$, and the integral can be conveniently calculated at $x \rightarrow
+ 0$ with $\psi $ from Eq.(\ref{5ea}).  The deflection angle $\Delta
\varphi \equiv \bar{\varphi}_{out}-\bar{\varphi}_{in}$ is $\Delta
\varphi = { \Delta p_{y}/ p }$ where $\Delta p_{y}= \langle
\hat{p_{y}} \rangle_{out} - \langle \hat{p_{y}} \rangle_{in} $. After
a simple calculation, \footnote{ Some caution is needed because of the
discontinuity of $\psi(+0,y)$ in Eq.(\ref{5ea}) at $y=0$.  The correct
limiting procedure, where details of the behaviour at $y=0$ are
unimportant, is the following: $\partial \psi /\partial y $ is
substituted by $\left(\psi (+0 , y+a)- \psi (+0,y -a) \right)/2a $ and
the limit $a \rightarrow 0$ is taken {\em after} the $y$-integration.}
\begin{equation}
\Delta p_{y} = - \hbar  |\psi_{\text{in}}(0)|_{N}^{2} \sin 2\pi
\tilde{\Phi} \ ,
\label{ida}
\end{equation}
where 
$ |\psi_{\text{in}}(0)|_{N}^{2} =  |\psi_{\text{in}}(0)|^{2}/
\left(\int_{-\infty }^{\infty }dy |\psi_{in}(y)|^{2}\right)$.

We see that indeed the AB-line deflects particles asymmetrically, with
the left-right asymmetry $\Delta \varphi = \Delta p_{y}/p $ controlled
by the parity-odd $\Phi $.  Unlike the Aharonov-Bohm effect ({\it
i.e.} the $\Phi$-dependence of the interference fringes), a
prerequisite for the deflection is a finite overlap of the {\it
incoming} wave, $\psi_{in}(0)\neq 0$, with the line.  By order of
magnitude $|\psi_{\text{in}}(0)|_{N}^{2}\sim 1/W$ and $\Delta p_{y}
\sim \hbar /W$, so that the deflection is of order of the beam angular
width $\varphi_{0} \sim \lambdabar /W$.

In Eq.(\ref{ida}), $\Delta p_{y}$ is the transverse momentum transfer
per collision. Multiplying it by the collision rate $\dot N $, one
gets a combination, ${\cal F}_{y}= \Delta p_{y} \dot N$, which has the
meaning of the force acting on the charge.  The collision rate is
found as $\dot N = \int_{-\infty }^{\infty }dy j_{x}(x,y)$, $j_{x}$
being the current density.  In the paraxial approximation $j_{x}= v
|\psi |^{2}$, and using Eq.(\ref{ida}) the force ${\cal F}_{y}= -
\hbar v |\psi_{in}(0) |^{2}$.  In terms of the full wave function
$\Psi_{in}(x,y)= e^{ikx}\psi_{in}(x,y)$ the effective ``Lorentz
force'' reads
\begin{equation}
 \bbox{{\cal F}}_{\perp} =  
\hbar \sin 2\pi \tilde{\Phi}\;
\bbox{J}_{in}
\bbox{\times e  }_{z}
\label{ifa}
\end{equation}
where $\bbox{J}_{in}$ is the current density in the {\it incident
wave} at the position of the line: $\bbox{J}_{in}= Re\; {\hbar \over i
m}\ \Psi^{*}_{in}\bbox{\nabla }\Psi_{in} |_{\bbox{r}=0}$.  Derived by
a different method and generalised to the case of {\it inhomogeneous}
$\bbox{J}_{in}(\bbox{r})$, this expression is in agreement with the
previous papers where a finite transverse force was found
\cite{OlaPop85,Son75,Son97,Cle68,NieHed95}.

If the flux in the line $\Phi \ll \Phi_{0}$ and $\hbar \sin 2\pi
\tilde{\Phi} \approx (e/c) \Phi$, Planck's constant disappears from
Eq.(\ref{ifa}), and the classical Lorentz force density is
recovered. As discussed in \cite{She98}, the transverse force in a
random {\it array} of the AB-lines amounts on average to the Lorentz
force in an effective magnetic field $B_{eff}= d_{AB} {\Phi_{0}\over
2\pi } \sin 2 \pi \tilde{\Phi}$, $d_{AB}$ being the density of the
lines in the array.  If $\Phi \ll \Phi_{0}$, the effective field
$B_{eff}$ equals to the magnetic induction $B=d_{AB}\Phi $.

A natural question to ask now is what may be wrong with Eq.(\ref{gma})
predicting zero force?  A short answer is that it is qualitatively
inadequate at small angles: As was previously noted by Berry {\it et
al.} \cite{BerChaLar80}, in the forward direction the outgoing wave
cannot be split into the incident and scattered pieces, and the
description based on the scattering amplitude looses its meaning.

In the scattering amplitude formalism, it is tacitly assumed that the
scattered waves generated by an infinite plane wave and a wide beam
differ insignificantly.  This is not true in the AB-case.  Here, the
plane incident wave solution does not have canonical asymptotics: As
known from the exact \cite{OlaPop85,BerChaLar80} (or paraxial)
solution, 
$\Psi = \cos \pi \tilde{\Phi}\exp (ikx)$
 in the forward
direction 
rather than
the assumed form $$\exp(ikx) + f \exp(ikr)/
\sqrt{ir} \, .$$  

On the contrary, a finite angular width incident beam
generates the solution with a {\it canonical} asymptotics form
\cite{LanLif3}, i.e.  $$ \Psi \rightarrow F_{in}(\pi -\varphi )
{e^{-ikr}\over \sqrt{r} } + F_{out}(\varphi ) {e^{ikr}\over \sqrt{r}
}.$$  Since (i) any physical incident wave is beam-like, and (ii) the
finiteness of the angular width turns out to be of qualitative
importance, a wave-packet-type theory operating with in/out amplitudes
$F_{in}, F_{out}$ is preferable.

Without separating into the scattered and unscattered pieces, the
outgoing amplitude can be generally related to the incoming one by the
$S$-matrix \cite{LanLif3}: $$F_{out}(\varphi)= \int_{-\infty}^{\infty}
d \varphi' S(\varphi , \varphi')F_{in}(\varphi').$$  In the paraxial
approximation, $F_{in,out}$ are defined via the momentum
representation: From $$\psi(x<0,y) = \int_{-\infty}^{\infty} d \varphi
F_{in}(\varphi ) e^{ik \varphi y - i {1\over{2}} kx \varphi^{2}}$$ one
finds $F_{in}$, whereas $F_{out}(\varphi)$ is extracted from $\psi
(x>0,y)$ in the same manner.  From Eq.(\ref{5ea}), $S(\varphi ,
\varphi')$ is the Fourier transform of $S(y,y')= \delta (y-y')\exp( -
\pi i \tilde{\Phi } \hat{y}) \cdot \exp(- \delta |y'|) $ where the last
factor with a regularization parameter $\delta = + 0$ is introduced
with the understanding that the incident wave has a finite extension
$\ll 1/ \delta $
in the $y$-direction.  Finally, the unitary $S$-matrix reads
\begin{equation}
S(\varphi , \varphi')= 
{1\over 2\pi i } {e^{-i \pi \tilde{\Phi}}\over \varphi-\varphi' - i\delta}
+ c.c.
\; \;\; , \;\; \delta = + 0 \; .
\label{yia}
\end{equation}
The overall phase of $S$-matrix is gauge-dependent; Eq.(\ref{yia})
agrees with a small angle limit of the $S-$matrix found in
\cite{Rui83} (see also \cite{Mor96}).  

Note that the ``scattering
amplitude'' formally defined as $\hat{f}= \hat{S}- \hat{1}$ does not
have the usual meaning because the ``scattering probability''
$|f(\varphi, \varphi ' )|^{2}$ \\(i) would be gauge dependent;\\ (ii)
would have intractable terms $\propto\left(\delta (\varphi - \varphi '
) \right)^{2}$.

For a reasonable $F_{in}(\varphi)$, the outgoing wave
$F_{out}(\varphi)= \hat{S}F_{in}(\varphi)$ is finite and continuous:
\begin{equation}
F_{out}(\varphi )= \cos \pi \tilde{\Phi}\,
F_{in}(\varphi )+ {1\over \pi}\sin \pi \tilde{\Phi}\,
\;{\sf P} \int 
d \varphi ' \,{F_{in}(\varphi')\over \varphi'
  -\varphi  }\, ,
\label{pja}
\end{equation}
here ${\sf P}$ stands for the principal value.  The first term on the
r.h.s.  gives the ``transmitted'' component, {\it i.e.}, the
attenuated incident wave, and the second one reproduces the
AB-scattering (it may be tagged as the ``scattered'' wave).  The
deflection of the beam as a whole can be found now as $\Delta \varphi
= \langle in |(\hat{S}^{\dagger}\varphi \hat{S} - \varphi )|in
\rangle$.  It is obvious from Eq.(\ref{pja}), that $ \Delta \varphi
\propto \sin 2\pi \tilde{\Phi}$ may originate only from the
interference of the transmitted and scattered waves.

The reason why Eq.(\ref{gma}) fails to give a finite force becomes
clear now.  The derivation which gives the force proportional to
$\sigma_{\perp}$ in Eq.(\ref{gma}), is based on the assumption that
the momentum distribution after collision can be found by adding the
corresponding probabilities for the incident and scattered waves.  In
the case of the AB-line, this classical assumption is erroneous: The
probability $|F_{out}|^{2}$ in Eq.(\ref{pja}), has also a contribution
from the interference of the transmitted and scattered wave.  The
parity-odd interference term proportional to $\sin 2\pi \tilde{\Phi}$
is the missing source of the left-right asymmetry in scattering and
the transverse force.  The interference takes place only in the almost
forward direction where $F_{in}(\varphi )\neq 0$.  Its importance is
peculiar to the AB-line case where the forward scattering is
anomalously strong.

As far as Eq.(\ref{lma}) is concerned, the critics in
\cite{ThoAoNiu97,WexTho98} (correct in the sense that the regular sum
in Eq.(\ref{lma}) cannot be identically derived from the singular at
$\varphi =0$ integral in Eq.(\ref{gma})) does not undermine
Eq.(\ref{lma}) since Eq.(\ref{gma}) is not a good starting point for
the derivation.  The expression for $\sigma_{\perp}$ in
Eq.(\ref{lma}), which is definitely valid in case of regular
scattering, does not show any alarming features in the AB-case (like
the singularity in Eq.(\ref{gma})) and remains applicable.  Its
validity can be confirmed by calculating the force as the flux of the
momentum flow tensor (as in \cite{Son75,Son97,OlaPop85}) through a
contour surrounding the AB-line.

In conclusion, the main concern of this paper has been to understand
how the finite transverse force can be reconciled with the symmetric
AB cross-section.  Using a new solution to the AB scattering problem
derived in the parabolic approximation Eqs.(\ref{4ea}-\ref{hfa}), the
transverse magnetic force has been ``observed'' in a gedanken
experiment where a beam-like incident wave hits the line and the
deflection of the beam Eq.(\ref{ida}) measures the magnetic force. For
a general incoming wave, the magnetic force is shown to be related to
the current in the {\it incident} wave at the position of the line
Eq.(\ref{ifa}).  It is argued that the $S$-matrix formalism
Eq.(\ref{yia}), rather than the scattering amplitude, is the adequate
language for the description of the forward AB-scattering.  It has
been shown that the expectation value of the momentum of the outgoing
particle and, therefore, the momentum transfer cannot be expressed via
the differential cross-section: peculiar to the AB-problem, the
interference of the scattered and the incident (transmitted) wave is
of qualitative importance.

Finally, a remark about the Iordanskii force in $He$-II, the
controversy about which has been the immediate motivation for the
paper.  The problem of the scattering of a phonon by the superflow
around a vortex line in $He$-II is similar to the AB-problem
\cite{Son97,WexTho98}.  The superflow is not a gauge-field, and the
equivalence of the two problems holds only in the lowest (linear)
approximation with respect to the vortex circulation $\kappa
\leftrightarrow \Phi $.  Then, the ``Lorentz force'' $\propto \Phi$ on
Eq.(\ref{ifa}) acting on the charge translates as (minus) the force
acting on the vortex line.  Integrating Eq.(\ref{ifa}) over the
phonons, one gets the Iordanskii force ($\propto\kappa $) transverse
to the normal flow acting on the vortex line from the normal
component.  The present calculations suggest that the Iordanskii force
is a close analog to the Lorentz force, and support the existing
understanding \cite{Ior65,Son75,Son97} of its role in vortex dynamics.

I am grateful to S. Levit, A. Mirlin, L. Pitaevskii, and P. W\"olfle
for discussions, and also to E. Sonin and J. Rammer for critical
remarks.  The study began during my stay at the Institut f\"ur Theorie
der Kondensierten Materie, Universit\"at Karlsruhe, and I would like
to thank all the members for their hospitality.  This work was
supported by SFB 195 der Deutsche Forschungsgemeinschaft and in part
by the Swedish Natural Science Research Council.

\end{document}